\documentclass[10pt]{article}

\usepackage[letterpaper]{geometry}
\usepackage{hicss}
\usepackage{times}
\usepackage[none]{hyphenat}
\usepackage{url}
\usepackage{latexsym}
\usepackage{tabularx}

\usepackage{indentfirst}
\usepackage{graphicx}
\graphicspath{{figures/}}
\usepackage[
    style=apa,
  ]{biblatex}
\usepackage{amsmath,amssymb,amsfonts}
\usepackage{paralist}
\usepackage{booktabs}
\usepackage{array}
\usepackage[bookmarks=false,hidelinks]{hyperref}
\usepackage[nameinlink]{cleveref}

\usepackage{enumitem}

\usepackage[bottom]{footmisc}

\title{DAG-Sword: A Simulator of Large-Scale Network Topologies for DAG-Oriented Proof-of-Work Blockchains}

\author{Martin Pere{\v{s}}{\'\i}ni\\
Brno University of Technology, FIT\\
{\underline{iperesini@fit.vutbr.cz}} \\ \\
\\ \\ \\\And
Tom{\'a}{\v{s}} Hladk{\'y}\\
Brno University of Technology, FIT\\
{\underline{xhladk15@stud.fit.vutbr.cz}} \\ \\
Ivan Homoliak\\
Brno University of Technology, FIT\\
{\underline{ihomoliak@fit.vutbr.cz}} \\ \And
Kamil Malinka\\
Brno University of Technology, FIT\\
{\underline{malinka@fit.vutbr.cz}} \\ \\
\\ \\ \\}

\date{}

\begin{document}
\maketitle

\begin{abstract}
	The blockchain brought interesting properties for many practical applications. 
	However, some properties, such as the transaction processing throughput remained limited, especially in Proof-of-Work blockchains.
	Therefore, several promising directions, such as sharding designs and DAG-based protocols emerged.
	In this paper, we focus on DAG-based consensus protocols and present a discrete-event simulator for them.
	Our simulator can simulate realistic blockchain networks created from data of a Bitcoin network, while its network configuration and topology can be customized.
	The simulated network consists of honest and malicious miners.
	Malicious miners do not make any attack on consensus itself. 
	Instead, they use a different transaction selection strategy than honest miners (who select transactions randomly) with the intention to earn unfairly more profits than honest miners at the cost of downgrading the protocol performance by duplicate transactions. 
	As a consequence, this harms the performance of some DAG-based protocols (e.g., PHANTOM and GHOSTDAG) in terms of transaction processing throughput, which we demonstrate in our experiments and extend the results of the related work that contains a small-scale network of 10 nodes by the results obtained on a large-scale network with 7000 nodes.
	Next, we empirically compare different algorithms for the mempool structure, and we propose a composite mempool structure that is memory-efficient and thus convenient for simulations of resource-demanding large-scale networks.
\end{abstract}

\section{Introduction}
\label{section:Introduction}
In contrast to traditional databases, blockchains provide interesting features such as decentralization, immutability, transparency, and censorship resistance. These features open new possibilities for a wide range of applications.
However, the limited blockchain throughput is one of the most significant problems preventing them from reaching their full potential. As a response to the throughput bottleneck of a traditional single-chain structure of blockchains, several approaches have been proposed in the literature involving sharding~\textcite{inproc:ZamaniM018-rapidchain}, \textcite{inproc:Kokoris-KogiasJ18-omniledger}, \textcite{inproc:luuElastico2016secure}, sidechains~\textcite{misc:poon2016bitcoinLightning}, \textcite{misc:poon2017plasma}, \textcite{misc:2020Polkadot}, and Direct Acyclic Graphs (DAGs)~\textcite{inproc:lewenberg2015inclusive}, \textcite{misc:sompolinsky2016spectre}, \textcite{inbook:phantomGhostdag}.
Many proposals contain empirical evaluation using custom simulation experiments; however, the results are often not reproducible since the authors do not provide the source code of their simulator and its settings.

This fact motivated our research, and we argue that besides security analysis and theoretical analysis of the proposed consensus designs, it is critically important to evaluate their properties reproducibly.
Nevertheless, this might be problematic as some of the modeled properties can only manifest in a network with a huge number of nodes. One option to plausibly evaluate the properties of the proposed consensus protocols is to deploy them in a real network and test their properties in such an environment. 
However, this option is extremely expensive because of the large number of nodes that execute expensive mining computations (e.g., the Bitcoin network has more than 15,000 consensus nodes\footnote{Captured from \url{https://bitnodes.io/nodes/} on 12/2022. }). Due to this reason, the simulation remains the only option to investigate blockchain properties in many cases. 
However, the design of a one-size-fits-all simulator is very challenging and heterogeneous, and therefore most simulators focus on specific consensus protocols and only some of their aspects in particular.
We follow this direction and focus on the niche of the simulators aiming at DAG-based consensus protocols.
Therefore, we propose an open-source simulator that focuses on DAG-based consensus protocols, and we make it \href{https://github.com/Tem12/DAG-simulator}{available}.\footnote{\url{https://github.com/Tem12/DAG-simulator}} 
In the context of this paper, we demonstrate the utility of our simulator on PHANTOM protocol and its optimization GHOSTDAG~\textcite{inbook:phantomGhostdag}.

\paragraph{Contributions.}
Our contributions are as follows:
\begin{enumerate}
	\item We propose a simulator that can simulate realistic network topology of DAG-based consensus protocols while capturing information on miners' profits and transaction throughput of the network.
	\item We introduce optimization into the modeling and implementation of the mempool, enabling us to efficiently accommodate different transaction selection strategies at the same simulation runs. 
	\item We demonstrate the simulator on two incentive strategies. One is honest and is in line with the incentive scheme of the consensus protocols. Another one enables us to study malicious behaviors as part of incentive-oriented attacks.
\end{enumerate}

\section{Simulator Design}
\label{section:Design}
We propose a simulator for DAG-oriented consensus protocols to simulate large-scale blockchain networks with topology resembling Bitcoin network.
Our simulator focuses on obtaining knowledge about miners' profits, network transaction throughput (i.e., the number of processed transactions per second), and the transaction collision rate (i.e., the number of transactions that are duplicates and thus do not contribute to the throughput). 
A miner that produces a block is rewarded from two sources: the block reward and transaction fees. The value of the block reward is deterministic and is known during mining. 
The reward from transaction fees is simply the sum of all the transaction fees in the block.
The most well-known DAG-protocols that aim at maximizing the transactional throughput (i.e., PHANTOM \& GHOSTDAG~\textcite{inbook:phantomGhostdag}, and Spectre~\textcite{misc:sompolinsky2016spectre}) propose to select transactions into block randomly\footnote{Not necessarily uniformly at random.} to avoid potential transaction collision in favor the throughput.
However, greedy miners in anonymous permissionless systems such as blockchain (without a possibility to punish misbehavior) might violate this rule and select transactions based on the fees (greedily) and thus attempt to maximize their profits.
Therefore, the incentive strategy of the protocol impacts the properties of DAG protocols, such as fair distribution of rewards and throughput.

To describe a blockchain simulator architecture, its functionality, and the security of supported protocols,  we adopt a multi-layer blockchain abstraction~\textcite{article:5layer_performance_evaluation}, \textcite{homoliak2020security}.
We focus on the network, consensus, and data layer; not the application layer.
Therefore, applications leveraging from the consensus protocols we experiment with are agnostic to our work.
The proposed simulator should fit the needs of simulating DAG-based protocols with realistic network sizes. Note that we do not focus on network attacks on blockchains such as DoS on connectivity or local resources, eclipse attacks, routing attacks, or DNS attacks~\textcite{homoliak2020security}. 
Thus, we can abstract from most of the details in the consensus layer and its properties, such as Merkle tree aggregation, SHA256 hash algorithm, signature verification, block verification, etc.
Next, we abstract from the ordering of blocks within the DAG structure of considered protocols since the transaction selection strategies we are modeling are agnostic to them.

\subsection*{\textbf{Incentive Strategies}}
In the context of this work, we implement two transaction selection strategies:
\begin{itemize}
	\item \textbf{Honest (random) selection} -- miners are selecting the transaction randomly, which is in line with the DAG-based protocols we simulate in this work~\cite{inbook:phantomGhostdag}.
	\item \textbf{Malicious (max-fee) strategy} -- miners are selecting transactions based on the highest fees, maximizing their profit, and violating the assumptions of the simulated protocols. 
\end{itemize}

\subsection*{\textbf{Functional Requirements}}
The simulator that we are designing should provide the following features:
\begin{itemize}[noitemsep,topsep=1pt,parsep=1pt,partopsep=1pt]
	\item Utilizes a discrete event simulation model.
	\item Includes a network layer that incorporates block propagation, allowing adjustment of latency between peers, and facilitates the customization of network topology.
	\item Features a data layer for transactions, encompassing various fees and sizes.
	\item Employs an optimized mempool data structure that efficiently handles different transaction selection strategies.
	\item Implements a rational selection strategy inspired by Bitcoin, where miners prioritize transactions with higher fees to maximize their profits.
	\item Implements a random selection strategy proposed by PHANTOM/GHOSTDAG, randomly selecting transactions to reduce the occurrence of transaction collisions.
\end{itemize}

\medskip
These features align with the related work~\textcite{misc:dagSimulatorProposal} that we are extending.
The authors of this work analyze different incentive mechanisms with a combination of different transaction selection strategies. Our simulator aims to analyze different incentives and miners' behavior under more realistic conditions (e.g., network size and latency similar to Bitcoin).

In this work, we conducted several experiments to analyze the simulator and its capabilities to model these attacks with different parameters.
One such experiment is detailed in \Cref{section:Evaluation} and shows how one malicious miner makes significantly more profit than other miners in a realistic Bitcoin-like network.
The experiment further confirms that malicious actors have a negative impact on transaction throughput.

\subsection{Simulator Base}
\label{subsection:simulatorBase}
When we designed the simulator, we chose to build on the existing simulator created by~\textcite{misc:sim_bitcoin_mining_simulator:andresen}, which is based on a discrete event simulation model and implemented in C\texttt{++}.
This simulator aims to verify miners' block rewards according to their mining power and stale block rate. In the original simulator, there is no inclusion of transactions in blocks.
The examples in Andresen's simulator include network topologies with ten or fewer miners connected randomly or in a circular topology.
Results are analyzed from collected information at the end of the simulation and show how network topology and latency parameters affect miners' profit. This simulator contains two different events that occur during the simulation:

\paragraph{Block generation.}
Miners selected by their mining power generate blocks continuously in time defined by the following probability distribution: 
\begin{eqnarray}
	f_{\mathbb{T}}(t)=\Lambda e^{-\mathbf{t}\Lambda},   
\end{eqnarray}
where $\Lambda=\frac{1}{\lambda}$, $e$ is the Euler's number, and $\lambda=600$ seconds is the average time to create a block in Bitcoin~\textcite{article:block_arrival}, \textcite{misc:nakamotoBtc}.

\paragraph{Block propagation.}
Miners receive blocks from other miners through a peer-to-peer network with a delay specified in the configuration file for each connection.

\subsection*{\textbf{Added Events}}
To be able to simulate miners that utilize mempools and transactions, we have added two additional events:

\begin{itemize}[noitemsep,topsep=1pt,parsep=1pt,partopsep=1pt]
	\item \textbf{Initial transaction generation}. Generate a specified number of transactions for each miner at the start of the simulation. At the moment, once this event is done, the contents of miners' mempools are the same.
	\item \textbf{Continuous transaction generation}.
	$x$ is the number of transactions to generate, and $y$ is the delay between the next transaction generation. These values are then in the following interval: $n\leq
	\mathbf{x}\leq
	m$ and $p\leq
	\mathbf{y}\leq
	q$, where $n,m,p,q$ are chosen parameters before the simulation starts. Both intervals follow the uniform distribution; however, the fixed values are also supported.
\end{itemize}

\subsection{Simulator Structure}
The structure of the simulator can be divided into three main parts, as shown in \Cref{fig:simBlackBox}:
\begin{figure}[t]
	\centering
	\includegraphics[width=0.9\linewidth]{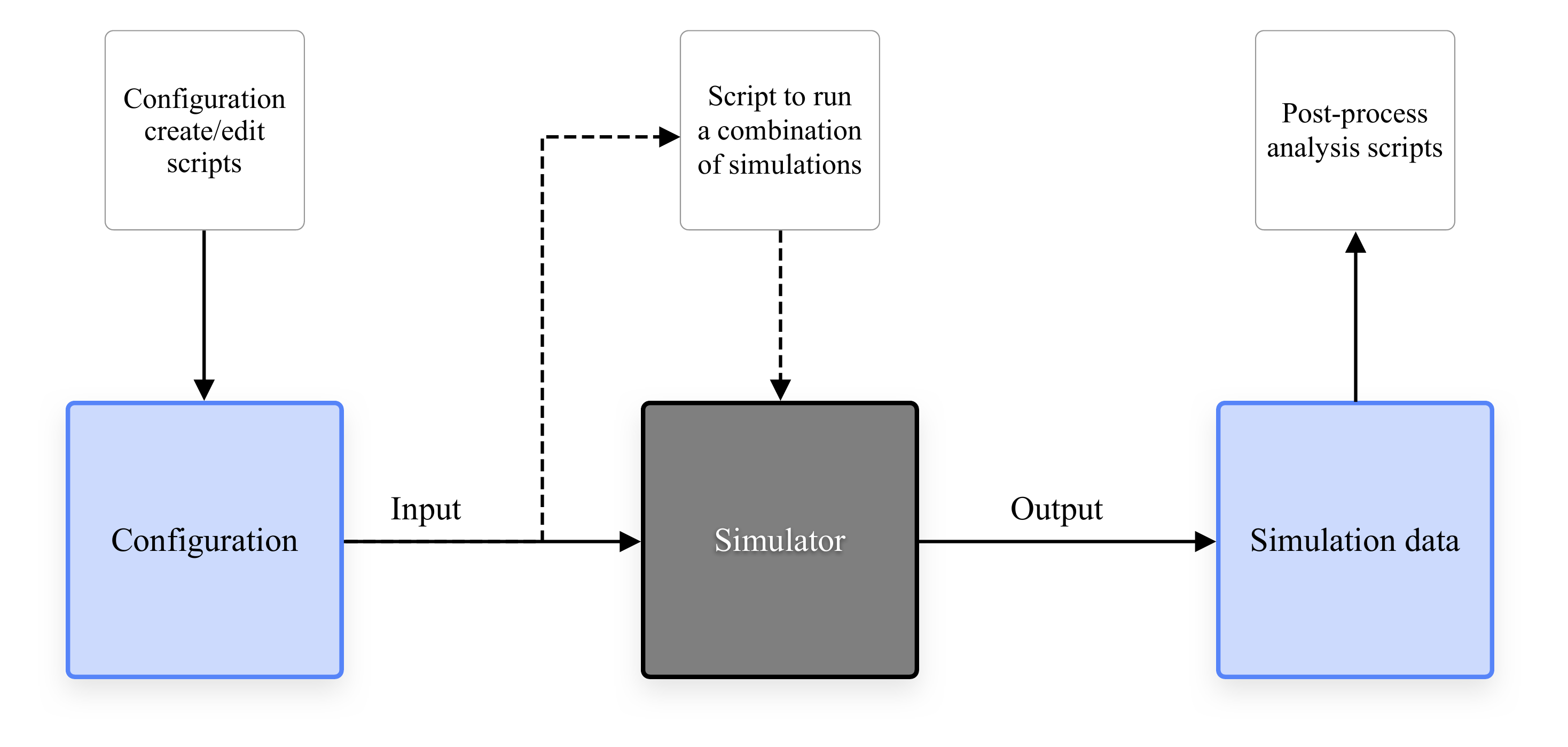}
	\caption{The conceptual structure of our simulator.}
	\label{fig:simBlackBox}
	\vspace{-0.5cm}
\end{figure}

\subsection*{1) Configuration Scripts}
To create a realistic blockchain network, we prepared scripts that follow a discrete distribution for the number of consensus nodes (i.e., miners) and the block propagation delay using data collected from the real Bitcoin network (see details below).

\paragraph{Realistic network.}
\label{subsection:realisticBlockchainNetwork}
Our proposal enables us to model a realistic blockchain network by using data from a Bitcoin network to verify the properties of DAG consensus protocols.
While we acknowledge that an Ethereum network could also serve as a baseline for large-scale networks, we argue that the DAG-oriented protocols are primarily an extension of Proof-of-Work, and thus, the Bitcoin-like network is the most suitable choice for DAG-based simulators.
The simulator itself also includes created network configurations from these data of the size of 7,592 nodes.\footnote{Number 7,592 comes from \url{https://bitnodes.io/nodes/}, and it states the number of IPv4 reachable nodes in the Bitcoin network on November 24th, 2021, when the first topology was created.}
Furthermore, the configuration script allows us to specify custom discrete distributions and create a network similar to other blockchains (e.g., Ethereum) or customized configuration for experimental purposes.

\begin{figure*}[t]
	\centering
	\vspace{-0.6cm}
	\begin{minipage}[t]{.47\textwidth}
		\centering
		\includegraphics[width=0.9\textwidth]{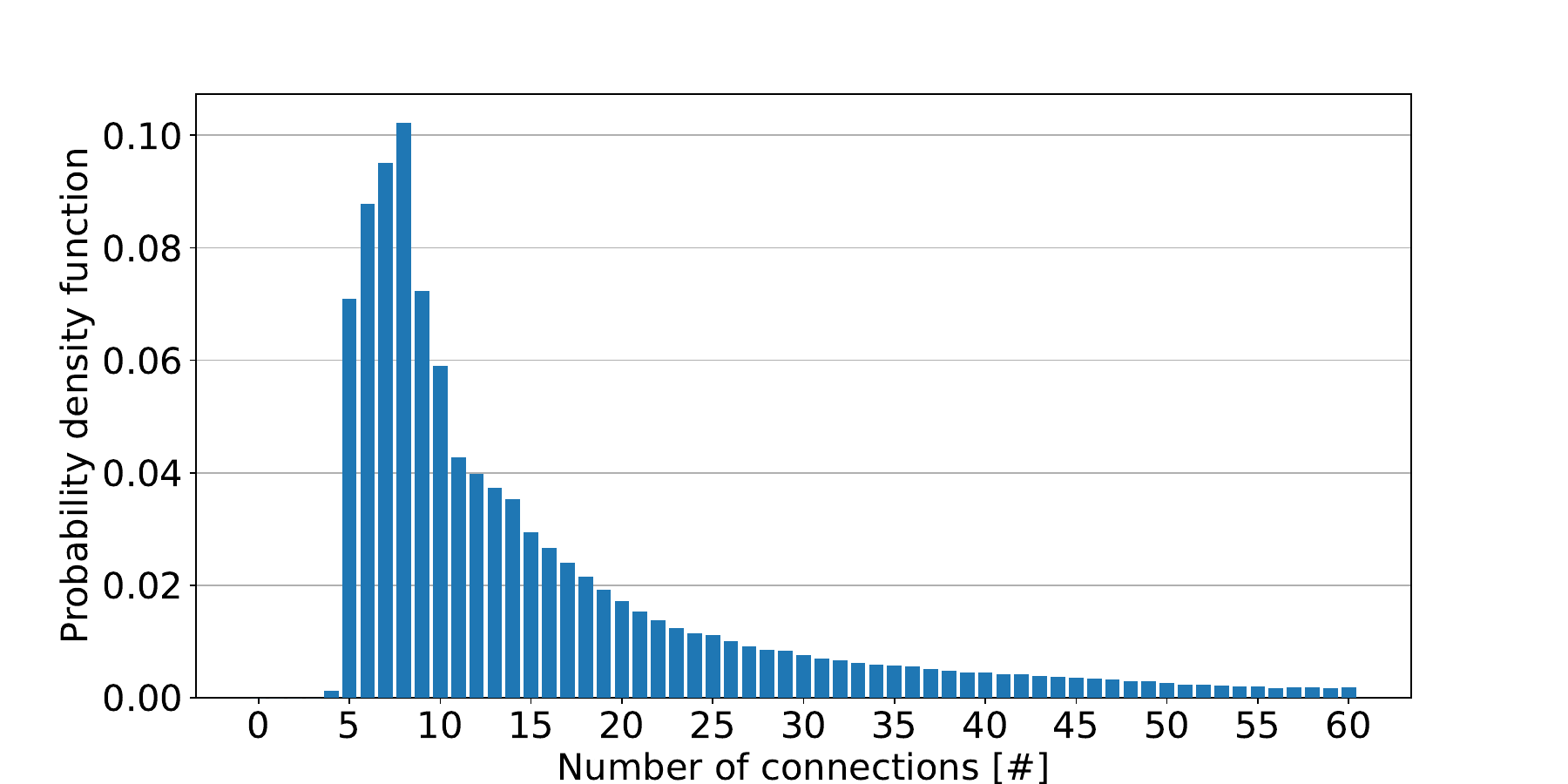}
		\caption{Distribution of the number of peer connections per node (i.e., node degree), created from data published in~\textcite{article:btcDeliveryNetwork}.}
		\label{fig:networkNumberOfConnection}
	\end{minipage}\hfill
	\begin{minipage}[t]{.48\textwidth}
		\centering
		\includegraphics[width=0.9\textwidth]{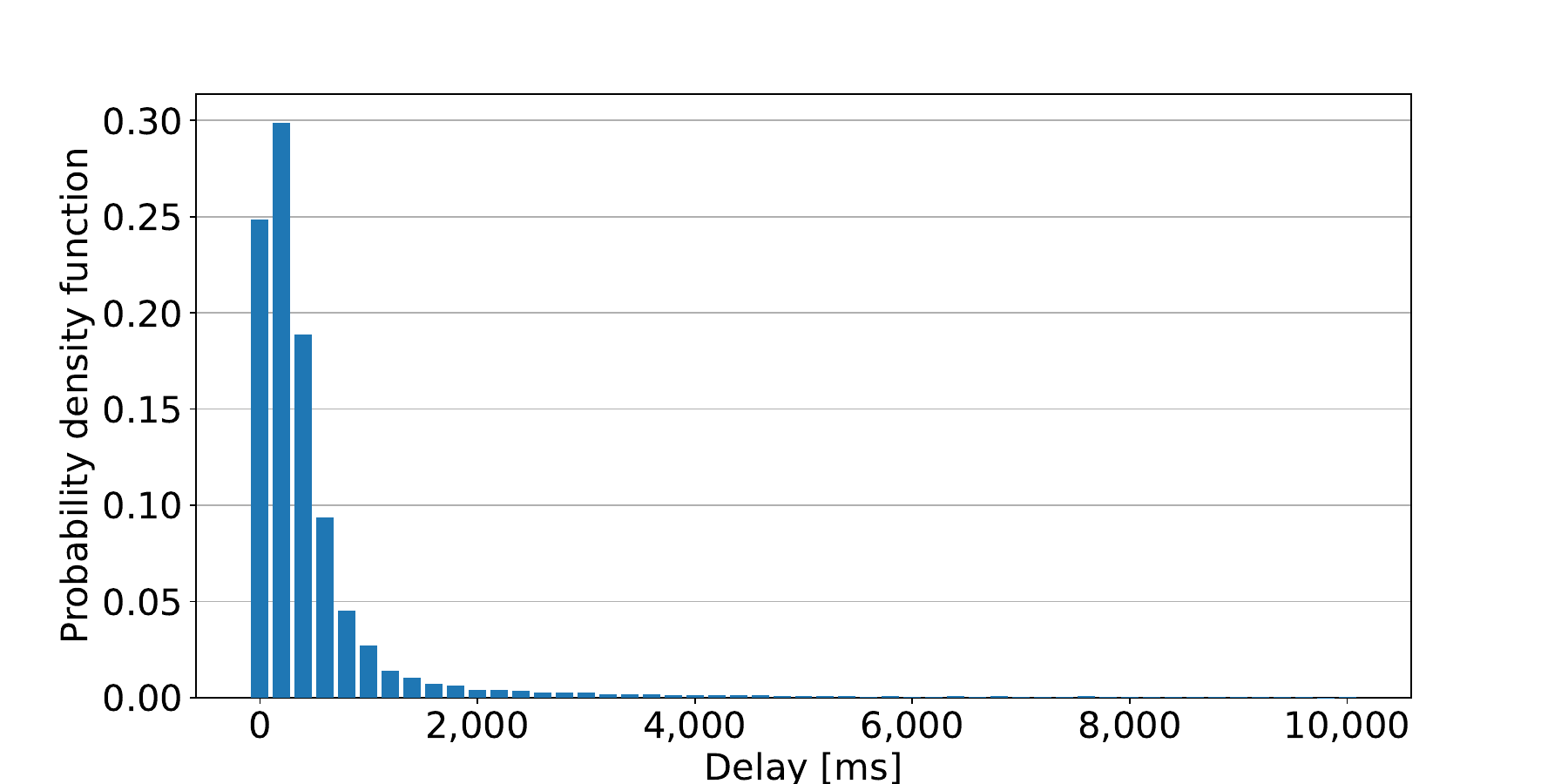}
		
		\caption{Distribution of the network propagation delay (for blocks), created from public data~\textcite{misc:btcMonitoringWebsite} on 2023-02-28. 
		}
		\vspace{-0.5cm}    
		\label{fig:networkBlockPropagationLatency}
	\end{minipage}
\end{figure*}

\paragraph{Number of connections per node.}
Nodes in the Bitcoin network have a different number of peer connections.
We created a discrete distribution from data published in~\textcite{article:btcDeliveryNetwork}. The distribution of nodes is shown in \Cref{fig:networkNumberOfConnection}. 
The parameters of this distribution are specified in the configuration file.
\paragraph{Block propagation delay.}
We created a discrete distribution to simulate block propagation delay between nodes. It was parametrized to fit a running node in the Bitcoin network from 24-hour data published in~\textcite{misc:btcMonitoringWebsite} on February 28th, 2022. \Cref{fig:networkBlockPropagationLatency} shows a portion of the distribution in which we can see that most blocks were propagated in under two seconds. Besides, the propagation delay depends on the block size. As stated in \Cref{subsection:simulatorBase}, our simulation requires a fixed number of transactions in each block. Thus, the block size and the propagation time of different blocks are the same. 

\subsection*{2) Simulator}
To run multiple simulations in parallel, we abandoned the original real-time data processing that imposed linear memory complexity due to storing data for each block to calculate results when the simulation ends. 
Instead, we propose a post-processing method to gather all data during a simulation and process them afterward. This approach allows us to analyze gathered data multiple times (e.g., applying different block rewards) and not lose any information even when the simulation crashes or does not finish. 
The simulator with the post-processing method generates the following types of data contained in different files.

\paragraph{Progress File.}
The file contains simulation information in a human-readable format and further continues with progress information. Once the simulation ends, it also prints its total duration. Suppose an error occurs when the miner does not have enough transactions to generate a block. In such a case, the simulation exits sooner than expected and prints an error message, taking a snapshot of all miners' mempool content.
The content of this file is also mirrored to the standard output. 

\paragraph{Data File.}
This file contains all the details about the generated blocks. When the block is mined, its information is printed in this file before propagation. Block information contains included \textit{transaction id}, \textit{transaction fee}, \textit{block id}, \textit{height} (depth), and \textit{miner id} who created it. Miner id is an index loaded from the configuration for each miner at the start of the simulation. The content of this file is stored in the CSV format. Post-processing scripts for collision and profit analysis, then load this file.

\paragraph{Metadata File.}
The purpose of this file is to facilitate post-processing scripts for the user by providing a formatted version of the data that is printed at the start of the simulation (in progress).
The data contained in this file is similar to what is printed during the simulation, but it is arranged in a way that makes it easier to parse for the post-processing parser method.

\subsection*{3) Post-processing Analysis Scripts}
Output data are analyzed for miners' profits and transaction duplicates, each within its specific script: 
\paragraph{Collision analysis script.}
This script requires only one argument, a path to a data file generated by simulation. It automatically parses the name and opens the metadata file to get all the required information. The produced output contains the number of unique transactions and transactions with two or more duplicates (up to five). It also outputs duplicate transactions along with the weighted sum of duplicated transactions and the duplication rate. 
\paragraph{Profit analysis script.}
It gets all the required information from metadata. An additional required argument is the mining power threshold. This argument specifies the minimum mining power required for the miner to be analyzed separately. Miners below the threshold will be merged during the analysis. As an optional argument, there can be a specified reward per block.
\label{section:Implementation}
The simulator is executed with a configuration describing the blockchain network, specifying miners' connections with block propagation latency and mining power. Program parameters specify further options.

\section{Mempool Optimizations}
\label{section:Optimizations}
The simulation spends most of its time on basic mempool operations such as sorting, inserting, and deleting. For this reason, the implementation of mempool is essential.
To efficiently simulate a complex network, we require optimizations that accelerate the simulation and minimize the memory utilization, enabling us to run more simulation processes simultaneously. 

Our post-processing method achieves lower memory usage than original real-time processing. However, to further speed up the simulation process, we focus on optimizing the mempool data structure because a simulation spends most of the time accessing the mempool.
We compared several candidate data structures (see \autoref{tab:mempoolStructuresAccessComparison} for mempool that will be described in the following.

\begin{table}[b!]
	\vspace{-0.4cm}
	\scriptsize
	\renewcommand{\arraystretch}{1.5}
	\caption{A comparison of time complexities of hashtable, red-black tree, and their combination.}
	
	\centering
	
	\begin{tabular}{l|c|c|c}
		\toprule
		\cmidrule(r){1-2}
		Access method & Hashtable & Red-black tree & A combination \\
		\midrule \hline
		Direct & $O(1)$ & $O(log(n))$ & $O(1)$ \\
		\hline
		Random & $O(\frac{m}{n})$ & $O(n)$ & $O(\frac{m}{n})$ \\
		\hline
		Sorted & $O(n*log(n))$ & $O(log(n))$ & $O(log(n))$ \\
		\bottomrule
	\end{tabular}
	\label{tab:mempoolStructuresAccessComparison}
\end{table}

\subsection{Properties of candidate data structures for the mempool}
\label{subsection:comparisonOfDataStructures}
\begin{itemize}
	
	\item \textit{\textbf{Hashtable.}} It has an average time complexity of $O(1)$ for search, insert, and remove operations.
	However, the downside is that it requires sorting all elements when accessing transactions in ascending or descending order, which is too expensive for the use case of the mempool that for some incentive strategies might require ordering all transactions in the mempool. 	
	\item \textit{\textbf{Red-black tree.}} It is an efficient data structure with an average time complexity of $O(log(n))$ for search, insert, and removal operations.
	Compared to the hashtable, it implicitly sorts all elements by a key, which imposes certain overhead during modifications, which are, however, very frequent in the use case of the mempool.
	
	\item \textit{\textbf{AVL tree.}} The time complexities for search, insert, and removal operations of the AVL tree are equivalent to those of the red-black tree, each exhibiting an $O(log(n))$ performance. However, compared to red-black trees, AVL trees are faster due to their more strict balancing of elements for lookup-intensive applications. On the other hand, AVL Trees might be slower in insert and removal operations because they require a slightly larger number of rotations to maintain the elements' balance.
	While fast search is important for a mempool structure, the red-black tree is a better choice overall due to its faster insert and removal times and its ability to perform other operations efficiently since mempool requires very frequent modifications.
	
	\item \textit{\textbf{Binary (min/max) heap.}} It has the average time complexity of $O(1)$ for searching the minimum or maximum value, depending on whether it is a min-heap or a max-heap.
	However, it has the average time complexity of $O(n)$ for accessing the other values.
	Implementing both a min-heap and a max-heap and mirroring their contents can alleviate this issue. 
	Note that the heap and the red-black tree have problematic random access in the context of a mempool structure, making both of them less efficient for accessing the mempool.
	
\end{itemize}
Since we require efficient access for all three operations (search, insert, remove), the hashtable and red-black tree are the best candidates for the mempool.

\subsection{Classification of mempool operations}
We can assign mempool operations to the previously mentioned simulation events, as follows:

\paragraph{Initial transaction generation event}
\begin{itemize}
	\item \textit{Direct insert.} Initial transactions are generated with a unique id and inserted into each miner's mempool, enabling her to start mining.
\end{itemize}

\paragraph{Transaction generation event}
\begin{itemize}[noitemsep,topsep=1pt,parsep=1pt,partopsep=1pt]
	\item \textit{Sorted access and remove (ascending).} If the miner's mempool is full (i.e., filled), she has to remove transactions.
	Because a malicious miner wants to earn the most profit, she removes transactions with the lowest fee.
	By default, this way is also preferred by honest miners.
	For a particular situation, when the fee of the newly generated transaction is lower than the fee of each currently stored transaction, the required amount of transactions is removed to make space for new ones.
	If this situation occurs in reality, malicious miners will keep older transactions with higher fees that have not been included in any block yet.
	With this approach, we achieve balanced transaction removal on full mempool capacity for both honest and malicious miners.
	\item \textit{Direct insert.} Each transaction is generated with a unique id and inserted into each miner's mempool.
\end{itemize}

\paragraph{Block generation event}
\begin{itemize}[noitemsep,topsep=1pt,parsep=1pt,partopsep=1pt]
	\item \textit{Sorted access (descending).} 
	This is required in rational selection strategy by a malicious miner, when she generates a block she also needs to sort (or has already sorted) a set of transactions from her mempool.
	\item \textit{Random access.} 
	Honest miners need to randomly access a transaction set to create a block while using a random selection strategy.
	\item \textit{Direct remove.} When an honest or malicious miner creates a block, she must remove included transactions from her mempool.
\end{itemize}

\paragraph{Block propagation event}
\begin{itemize}[noitemsep,topsep=1pt,parsep=1pt,partopsep=1pt]
	\item \textit{Direct remove.} When a mined block is propagated to other miners, they remove already included transactions from their mempool.
\end{itemize}

\begin{figure}[t]
	\centering
	\includegraphics[width=0.9\linewidth]{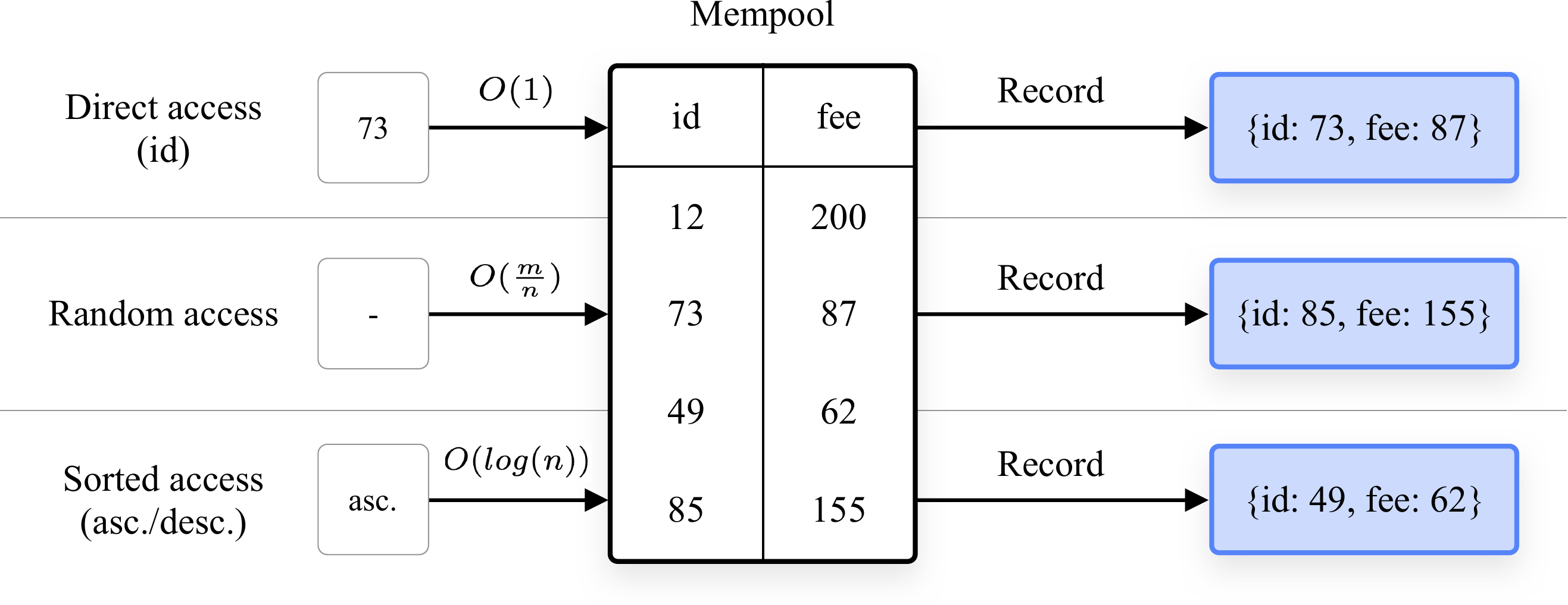}
	\caption{
		Various mempool access methods.             
	}
	\label{fig:mempoolDataStructure}
\end{figure}

\noindent
By examining existing operations of the mempool, we end up with three different access methods to the mempool (see \Cref{fig:mempoolDataStructure}) -- direct access, random access, and sorted access. Direct access returns a transaction requested by id. 
Random access returns a transaction randomly selected from the mempool. 
Sorted access returns one or more transactions in either ascending or descending order sorted by a fee.

\subsection{Random access to the mempool}
\label{subsection:mempoolRandomAccess}
It is important to note that we cannot approximate random access since it would skew the results (e.g., randomly choosing an entry by entry in a red-black
tree to access the element). Our simulations require that an honest miner's transaction selection is independent of the selection of other honest miners. Another important note is that random selection will be used most of the time in simulations because the assumption for malicious miners is to have less than 50\% mining power in total (i.e., more than 50\% involves 51\% attack). Therefore, we need an efficient pseudo-random selection algorithm for the mempool data structure.

For efficient access, we propose an algorithm \textit{random access} that can randomly select an element stored in the hashtable.
This algorithm randomly selects an initial index for a hashtable. If the bucket on this index contains an element, it is selected. 
Otherwise,  elements on indices above and below are checked. This process repeats until some element is found or until all buckets are investigated. If the bucket selection above or below exceeds the boundary of the hashtable, it is moved to the other side and continues in the same direction. In the case that a single bucket contains multiple elements, one is selected from a linked list with the time complexity of $O(n)$, where $n$ is the number of elements in the bucket. The average time complexity of this algorithm is $O(\frac{m}{n})$, where $m$ is hashtable capacity, and $n$ is the number of elements stored in a hashtable. The worst-case time complexity is $O(n)$.
For further comparison, we present the following algorithm modifications.

\paragraph{Begin.}
Instead of choosing the index randomly, it always starts with a zero value. Then, the value is incremented until it finds an element. After that, it randomly selects an element from a linked list if multiple elements are in the bucket. This approach gives skewed results because elements at the end of the hashtable will be selected rarely or never, depending on the fullness of the hashtable.

\paragraph{Equal key.}
This approach is comparable to \textit{begin} but does not involve utilizing a miner-specific (random) hashtable key.
The key content consists only of transactions; thus, all miners will have transactions on the same hashtable indices. 
For this reason, we expect this to lead to an enormous increase in transaction duplicities.

\begin{figure}[t]
	\vspace{-0.9cm}    
	\centering
	\includegraphics[width=0.98\linewidth]{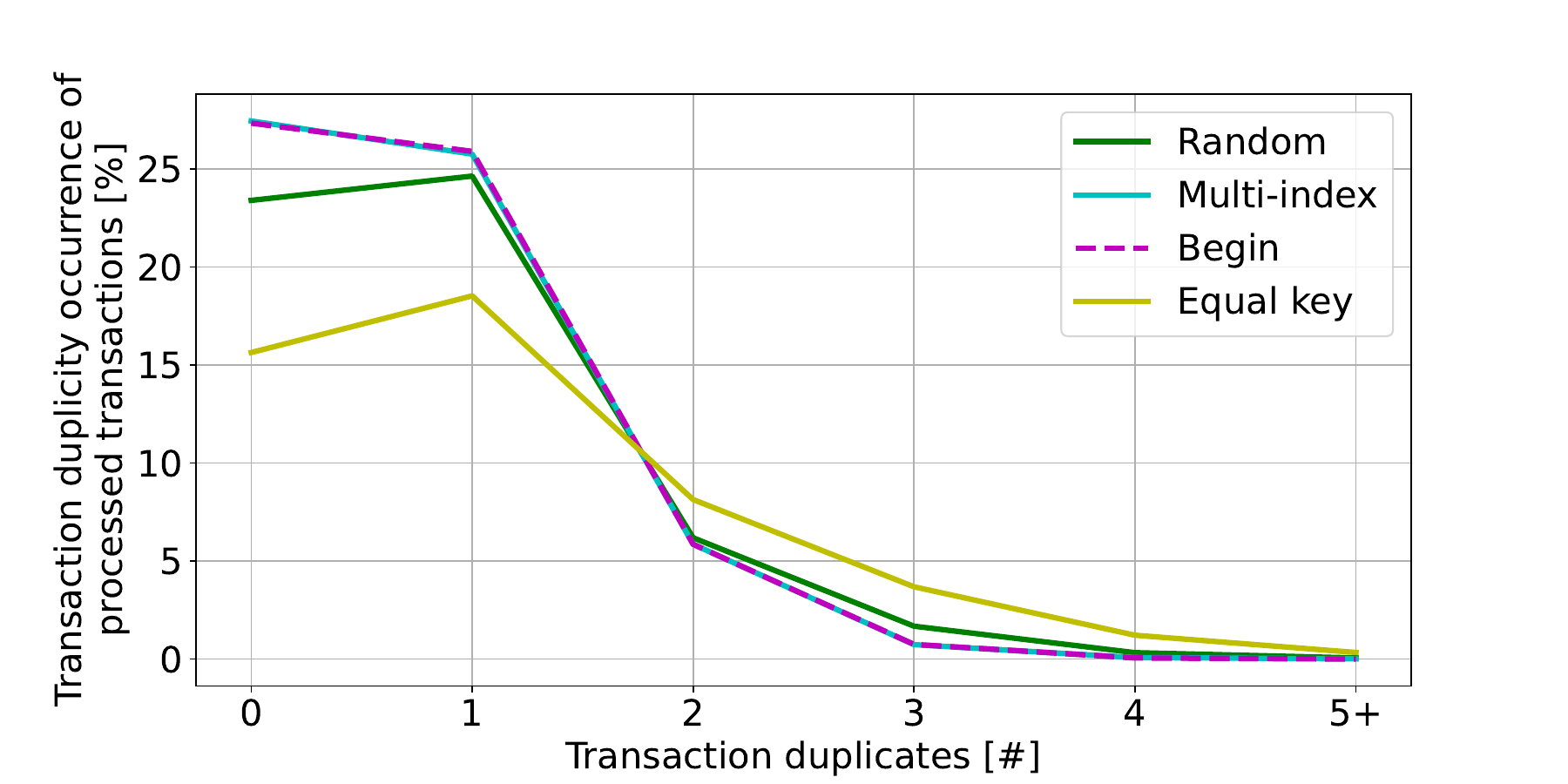}
	\caption{
		Transaction collision rate of different random selection algorithms.
	}
	\label{fig:randomGenerationExperimentResults}
	\vspace{-0.3cm}	
\end{figure}

\medskip
In addition, we also compare multi-index\footnote{\url{https://www.boost.org/doc/libs/1_78_0/libs/multi_index/doc}} implementation with previously mentioned algorithms. This was done by simulating a proposed realistic blockchain network with a 5-second block propagation delay (see \Cref{tab:randomGenerationExperimentParams} also containing other parameters). 
The fixed 5-second delay is much higher than the delay of a proposed blockchain network in \Cref{subsection:realisticBlockchainNetwork} because we want to achieve a higher transaction collision rate, which will be better reflected in the algorithm comparison results. Transaction collision rate also depends on parameters such as block size, mempool capacity, transaction generation size, transaction generation speed, and block creation rate. Results from this comparison can be seen in \Cref{fig:randomGenerationExperimentResults}.

\begin{table}[b]
	\vspace{-0.4cm}	
	\caption{Table of parameters used in simulations with different random selection algorithms.}
	\centering
	\scriptsize
	\begin{tabular}{l|l}
		\toprule
		\cmidrule(r){1-2}
		Parameter & Value \\
		\midrule
		Miner strategy & rational $|$ random \\
		\textbf{Malicious miners count} & 2 \\
		Honest miners count & 7,590 \\
		\textbf{Total malicious mining power} & 20\% \\
		Block creation time ($\lambda$) & 20 seconds \\
		Blocks & 1,000 \\
		Block propagation latency & 5 seconds \\
		\textbf{Block size} & 100 transactions \\
		Mempool capacity & 10,000 transactions \\
		Transaction generation & 60 to 160 seconds \\
		\bottomrule
	\end{tabular}
	\label{tab:randomGenerationExperimentParams}
\end{table}

\begin{table*}[t]
	\vspace{-0.7cm}
	\renewcommand{\arraystretch}{1.1}
	\tiny
	\centering
	\caption{Characteristics of selected simulators (evaluated on our best effort).}
	\begin{tabular}{ l|c|c|c|c|c|c|c  }
		\toprule
		Simulator & Scalability & Performance & Block Abstr. & Modularity & Post-Process & Network Topology & Event-Driven \\
		\midrule \hline
		BTCsim & $\times$ & \checkmark & $\times$ & \checkmark & - & $\times$ & \checkmark \\
		\hline
		Simbit & \checkmark & $\times$ & $\times$ & \checkmark & - & \checkmark & $\times$  \\
		\hline
		Shadow-Bitcoin & \checkmark & \checkmark & \checkmark (-) & - & \checkmark & \checkmark & $\times$ \\
		\hline
		Bitcoin-Simulator & - & - & \checkmark & \checkmark & \checkmark & \checkmark & \checkmark \\
		\hline
		BlockSim:Faria & $\times$ & \checkmark & \checkmark & - & \checkmark & - & \checkmark \\
		\hline
		SimBlock & \checkmark & - & \checkmark(-) & \checkmark(-) & - & \checkmark & \checkmark \\
		\hline
		\textbf{DAG-Sword} & \checkmark & \checkmark & \checkmark & \checkmark & \checkmark & \checkmark & \checkmark \\
		\hline
	\end{tabular}
	\label{tab:simulators_analysis_detailed}
	\vspace{-0.3cm}
\end{table*}

Results show that default (i.e., Boost implementation) \textit{multi-index} implementation behaves similarly as \textit{begin} modification -- the results vary only very little.
Besides that, \textit{equal key} has a much higher duplicity rate and a lower rate of total included transactions (i.e., throughput)
because there is not enough randomness of miner-specific id, which is used in \textit{begin} approach.
\textit{Random access} performance is not better than default \textit{multi-index} or \textit{begin} but produces more results that can be accurately replicated (with different simulation seeds).
Effective throughput stands for uniquely processed transaction rate.
Since \textit{random access} achieved the best results,
we will further utilize it for our next experiments.\footnote{It is worth noting that we are implementing a form of ``multi-index" but in a way that meets our specific needs.}

\begin{figure}[t]
	\vspace{-0.6cm}    
	\centering
	\includegraphics[width=0.9\linewidth]{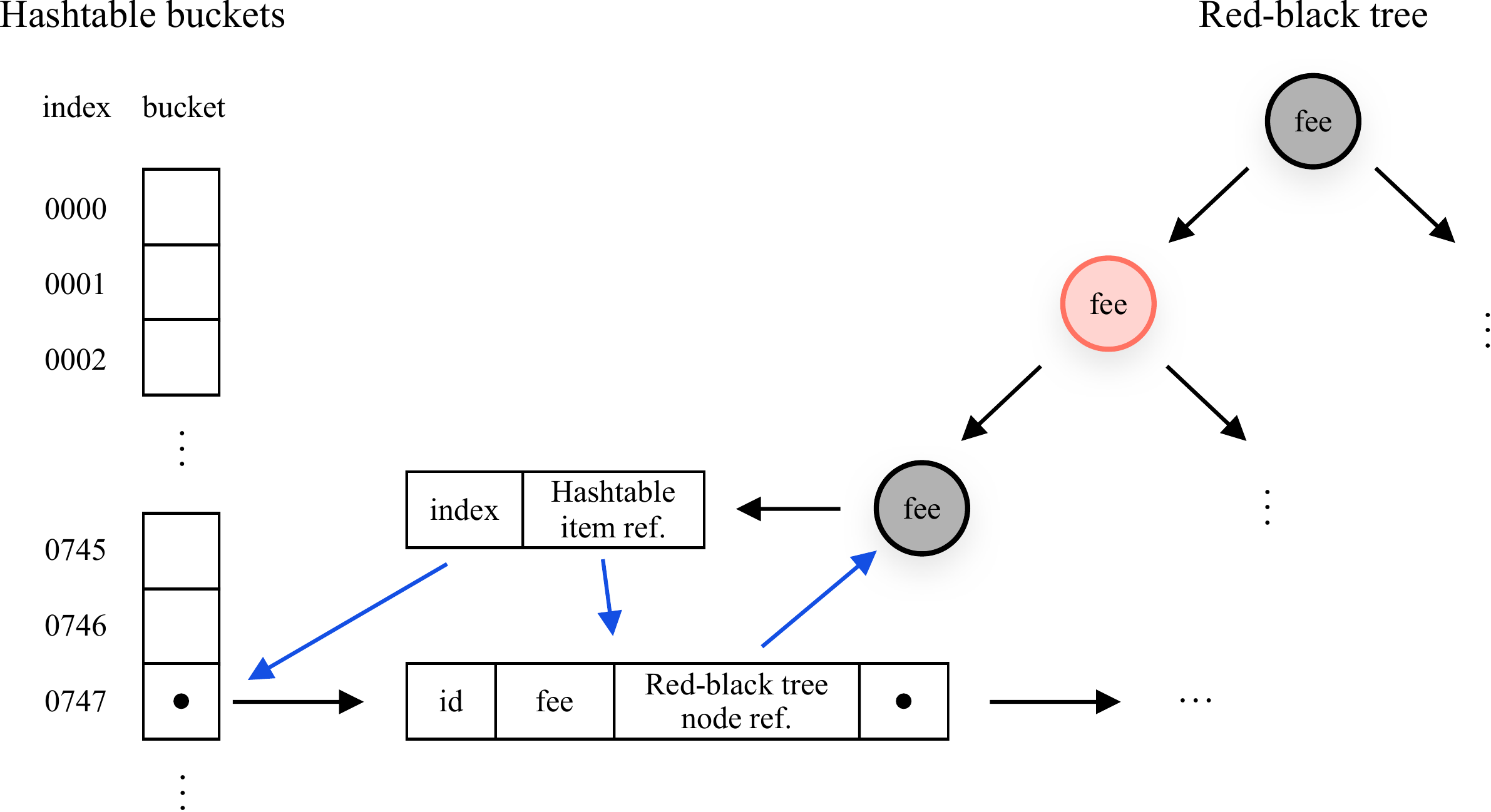}
	\caption{
		Our composite data structure -- a combination of the red-black tree and the hash table.
	}
	\label{fig:mempoolCombinedDataStructure}
	\vspace{-0.3cm}
\end{figure}

\subsection{Mempool Data Structure}
By analyzing hashtable and red-black data structures, we have compared their amortized time complexities for different types of accesses shown in \Cref{tab:mempoolStructuresAccessComparison}. Because both the hashtable and the red-black tree have advantages and disadvantages, we decided to create a unique data structure created by their combination. By choosing a simultaneous combination of these data structures, we benefit from both of them. A hashtable is used for direct and random accesses, and a red-black tree is used for sorted accesses (see \Cref{fig:mempoolCombinedDataStructure}). Such a composite structure has a higher memory usage (10.6 GB) compared to 6.5 GB of a hashtable in a network of 7,592 nodes, each with a full mempool of 10,000 transactions.
After simulating various mempool capacities, we compared our composite structure with a hashtable and found out that the composite structure outperforms it in all scenarios.

\section{Related Work}
\label{section:RelatedWork}
In this work, we extended the work of \cite{misc:dagSimulatorProposal}, who made similar experiments but on a topology consisting of only 10 nodes.  
In contrast, we employed the real network topology in our simulations as well as a few practical optimizations. 
Therefore, our results are more precise and plausible.

\paragraph{Simulators for Large-Scale Networks.}
The next lines provide an overview of existing simulators tailored for simulating large-scale blockchain networks (typically comprising more than 5,000 nodes).
We will delve into their applicability for simulating DAG-oriented blockchains, emphasizing aspects such as scalability (in terms of network size), performance (in terms of time spent in simulation run), modularity (extensibility and complexity to add new features), offline post-processing capabilities, and customization of network configuration.
\textcite{article:simulator_analysis_article} provides a summary of several simulators, out of which we selected 6 for further review and comparison. 

\textit{Simbit}~\textcite{misc:sim_Simbit} is a P2P network simulator implemented in JavaScript.
It offers significant flexibility in configuring network settings. However, it does not abstract blocks (transactions).
Instead, it employs a real-time simulation approach, relying on the concept of clients rather than event-driven simulation.
Nevertheless, it may encounter performance bottlenecks when simulating large-scale networks while providing excessive detail at the network layer.
\textit{Shadow-Bitcoin}~\textcite{misc:sim_shadow-bitcoin} is a simulator that integrates the actual Bitcoin implementation (Bitcoin Core).
This highly intricate simulator makes extending or customizing for specific purposes challenging, resulting in performance limitations.
It lacks a sufficient level of abstraction for swift prototyping.
\textit{Bitcoin-Simulator}~\textcite{inproc:paper_bitcoin-simulator} is built upon NS3\footnote{\url{https://www.nsnam.org/}}, and it completely models network transactions, even at the level of individual data frames.
While this detailed approach may potentially impact the overall simulation performance, the simulator compensates with its strong modularity and extensive network customization options, making it suitable for collecting data for subsequent statistical analysis.
\textit{BTCsim}~\textcite{misc:sim_BTCsim} is a basic Python simulator primarily designed to illustrate the concepts of a 51\% attack and selfish mining for security analysis purposes.
This simulator is quite simplistic, focusing solely on simulating blocks without transactions and functioning more as a mathematical model.
It does not offer support for network configuration or any form of customization.
\textit{BlockSim:Faria}~\textcite{article:paper_blocksim:faria} is a Python-based simulator designed to replicate the behavior of either the Bitcoin or Ethereum networks across three distinct locations: Ireland, Ohio, and Tokyo.
It encompasses a data layer and effectively simulates blocks with timestamp, difficulty, nonce, full transactions, and the previous block hash.
The simulator also includes an implementation of block mining difficulty adjustment and closely resembles a full node implementation rather than a high-level abstraction.
This simulator can be easily extended with new properties to simulate.
\textit{SimBlock}~\textcite{inproc:paper_simblock} written in Java offers relatively moderate extension simulator capabilities with a specific emphasis on the network and consensus layers.
The simulator provides comprehensive configuration options, enabling users to define various aspects of the simulation environment.
This includes configuring a network of regions with settings for download and upload speeds, latency, and regional distribution.
However, performance could potentially be a limiting factor.

None of the mentioned simulators explicitly incorporate DAG structure principles, but they possess the potential for adaptation through extensibility.
In contrast, DAG-Sword offers native support for DAGs and a moderate level of customizability for network configuration scripts, enabling the simulation of large networks. 

\section{Evaluation}
\label{section:Evaluation}
To verify the functionality of our simulator on DAG-based consensus protocols, we performed two experiments. 
The first experiment verified that the profit of a single malicious node should be higher than that of an honest node in a large network consisting of 7,592 nodes. The second experiment examined the profits of several malicious nodes with dominant mining power while gradually increasing their number.

\begin{figure}[t]
	\vspace{-0.7cm}	
	\centering
	\includegraphics[width=0.95\linewidth]{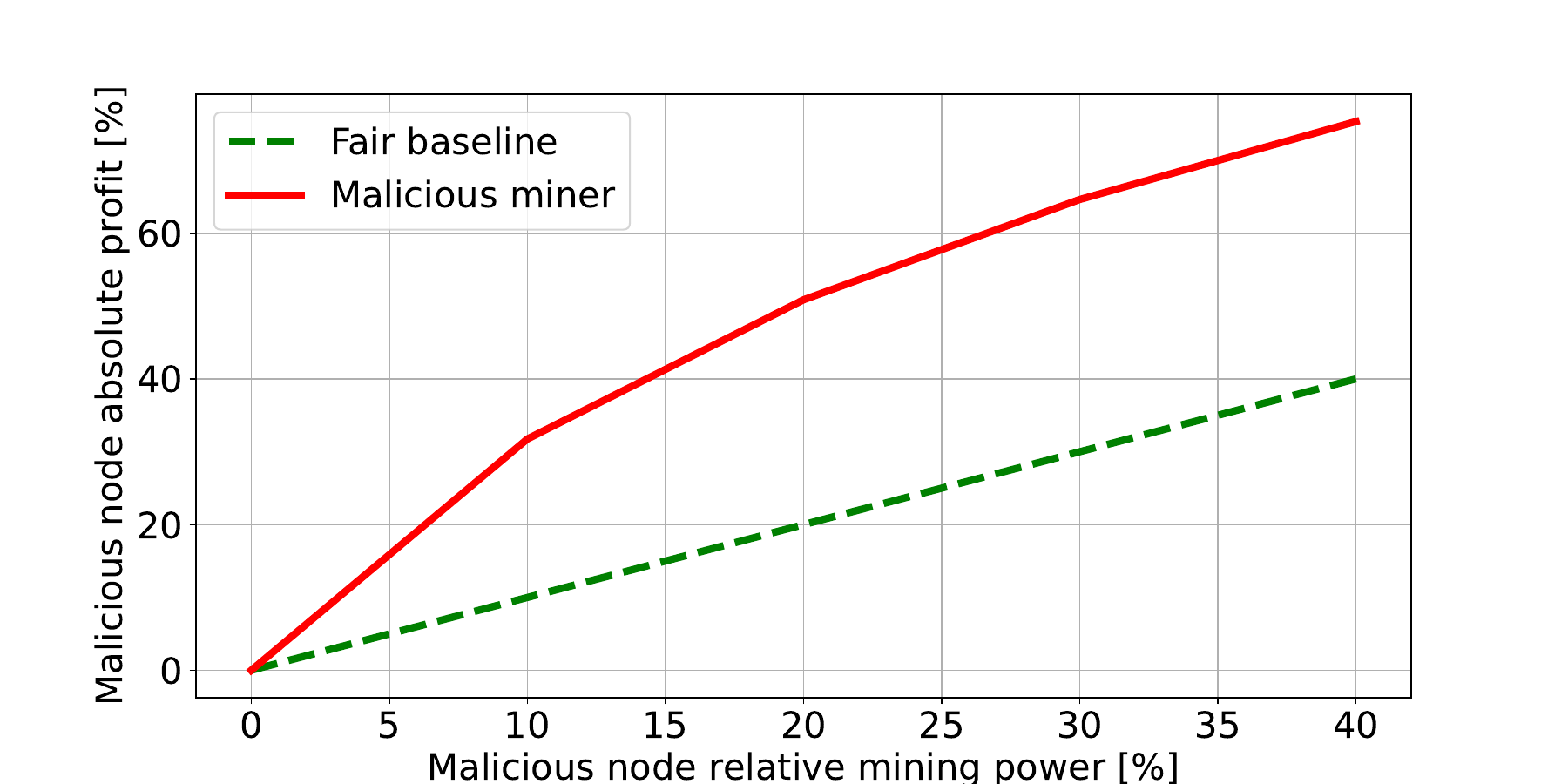}
	\caption{The profit of a malicious miner with varying mining power.}
	\label{fig:experiment1Results}
	\vspace{-0.4cm}	
\end{figure}

\paragraph{Experiment with a single malicious node.}
The experiment investigated whether a single malicious miner had an advantage in making profits compared to honest miners in a realistic network topology presented in \Cref{subsection:realisticBlockchainNetwork}.
We simulated a single malicious miner while continuously increasing her mining power relative to the network. 
Her connections to the other peers in a network might differ by assuming a well-connected node as opposed to a poorly connected node.
Therefore, our experiment involves different placements of the miner in the network. The results are averaged over multiple positions in the network. \Cref{fig:experiment1Results} shows the fair baseline, representing profits relative to the miner's mining power. Each miner should be rewarded commensurately for the number of resources spent in order to preserve fairness. However, the results show that malicious miner has significantly more profits. 

\begin{figure}[t]
	\vspace{-0.7cm}
	\centering
	\includegraphics[width=0.95\linewidth]{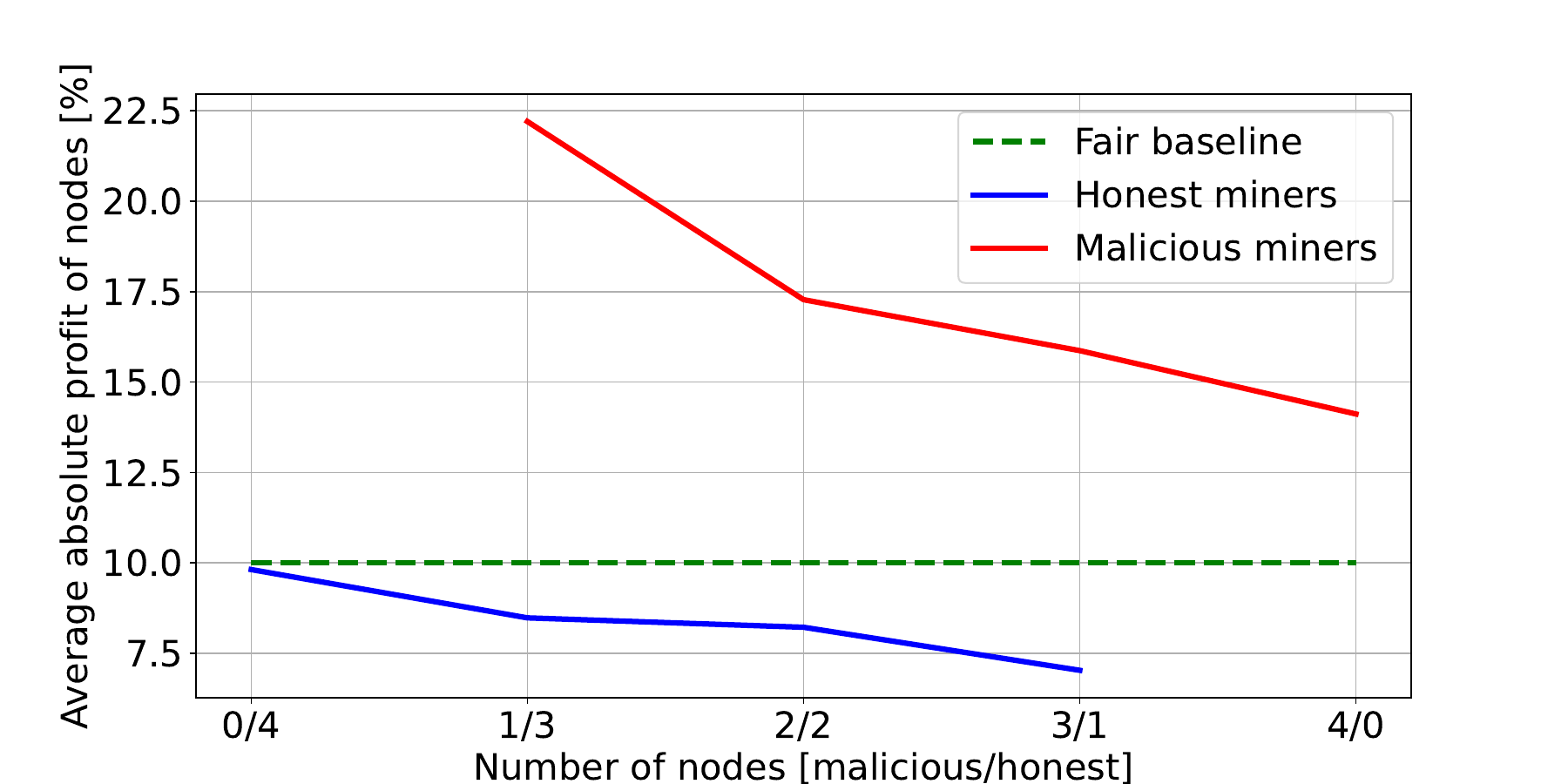}
	\caption{The averaged profits earned by honest and malicious nodes w.r.t. their mining power (each with 10\% of the total mining power).}
	\label{fig:experiment2Results}
	\vspace{-0.4cm}	
\end{figure}

\paragraph{Experiment with multiple malicious nodes.}
This experiment aimed to verify the profits of multiple malicious miners.
We simulated a network with 7,592 nodes, focusing on 4 nodes and assigning them mining power.
Each of these four nodes had 10\% of the mining power relative to the network.
The other nodes in the network were honest and assigned an equal portion of the remaining mining power.
The results are depicted in \Cref{fig:experiment2Results}, which shows that by increasing the number of malicious nodes, their profit decreases as there is greater competition, which is in line with the results from the previous work~\cite{misc:dagSimulatorProposal} that performed the experiments on a topology with only 10 nodes.

\section{Discussion}

\paragraph{Network Topology.}
While one may argue that we could consider other blockchain network topologies, such as the Ethereum network, the important point for us was to involve a realistic blockchain network, which was accomplished. Furthermore, we argue that the DAG-oriented protocols we investigated are primarily the extension of the Proof-of-Work protocol based on Bitcoin.
While this approach may not provide a direct analysis of DAG-based blockchains, we believe that given the lack of available data on DAG-oriented protocols, this approach is currently the most viable validation method.

\paragraph{Transaction Selection.}
We acknowledge that in Bitcoin, miners are incentivized to choose the highest-fee transactions.
However, we want to emphasize that our simulation aimed to analyze the incentives and behaviors of miners in the \textit{Bitcoin-like} network and extend these insights to DAG-oriented protocols that do not necessarily follow the highest-fee transaction selection strategy due to maximization of throughput.

\paragraph{Significance and Practicality of the Results.}
We repeated all experiments 10 times and depicted only the average value since the spread was minimal.
In the context of this work, we assumed that blocks in a DAG-based blockchain contain a fixed number of transactions, and there are no block rewards. However, in reality, there might be block rewards along with transaction fees, and blocks might contain a various number of transactions. 
Nevertheless, malicious miners are still incentivized to follow the random transaction selection strategy and profit more from such behavior.

\section{Conclusion}
\label{section:Conclusion}
In this work, we introduced a DAG-based blockchain simulator capable of accurately simulating network topologies similar in size to Bitcoin. Our analysis focused on the properties of PHANTOM/GHOSTDAG consensus protocols, comparing rational versus random transaction selection strategies. We confirmed the results from related work on a small proof-of-concept topology of 10 nodes are valid.
Efficient memory utilization and optimization of the simulator were prioritized, particularly regarding the mempool data structure. We assessed various implementation approaches in terms of performance. Our initial proof-of-concept attack demonstrated that malicious miners could achieve higher profits by employing a rational transaction strategy while reducing the overall network throughput.
It is worth noting that selecting appropriate input parameters for the simulator is a challenging task, as slight changes can yield significantly different outcomes. To further enhance and refine the simulator, we recommend implementing a consensus layer to investigate its behavior more comprehensively.
We also propose extending the simulator with a transaction propagation event to achieve more realistic results. Additionally, the flexibility of the simulator allows for adjusting different parameters, enabling the study of network behavior under diverse conditions.
In future work, we plan to experiment with a new transaction selection strategy, where miners would passively utilize their computational resources only occasionally.

\section{Acknowledgments}
This work was supported by the internal project of BUT (FIT-S-23-8151) and Cybersecurity Innovation Hub, Digital Europe Programme, no. 101083932. Computational resources were provided by the e-INFRA CZ project (ID:90254).

\printbibliography

\end{document}